\documentclass[aps,twocolumn,pre,showpacs]{revtex4}
\usepackage{graphics,amssymb,amsmath}

\begin{document}

\title{Growing Scale-Free Networks with Tunable Clustering}
\author {Petter \surname{Holme}}
\email{holme@tp.umu.se}
\author {Beom Jun \surname{Kim}}
\email{kim@tp.umu.se}
\affiliation {Department of Theoretical Physics,
  Ume{\aa} University, 901 87 Ume{\aa}, Sweden}

\begin{abstract}
We extend the standard scale-free network model to include
a ``triad formation step''.
We analyze the geometric properties of networks generated by this
algorithm both analytically and by numerical calculations, and
find that our model possesses the same characteristics as
the standard scale-free networks like the power-law degree
distribution and the small average geodesic length, but with the
high-clustering at the same time.
In our model, the clustering coefficient is also shown to
be tunable simply by changing a control parameter---the average
number of triad formation trials per time step.
\end{abstract}


\maketitle

  A great number of systems in many branches of science can be modeled as large
sparse graphs, sharing many geometrical properties~\cite{reviews}.  
For example: social
networks, computer networks, and metabolic networks of certain organisms all
have a logarithmically growing average geodesic (shortest path) length $\ell$
and an approximately algebraically decaying distribution of vertex degree.
In addition to this, social networks typically show a high clustering,
or local transitivity: If person $A$ knows $B$ and $C$, then $B$ and $C$ are
likely to know each other. 

Works on the geometry of social networks, which
is the main focus of the present paper, have originated from 
Rapoport's studies of disease spreading~\cite{rapoport}, and have been
further developed in Refs.~\cite{post_rapoport,FRO}. 
General mathematical models for random graphs with a structural 
bias are called the Markov graphs and were studied in Ref.~\cite{markov}.
In the physics literature,
 networks with high clustering are commonly modeled by the small-world network
model of Watts and Strogatz (WS)~\cite{WS}, while networks with
the power-law degree distribution by the scale-free network model 
of Barab{\'a}si and Albert (BA)~\cite{SF}. Although both models have a 
logarithmically increasing $\ell$ with the network size, each model
lacks the property of the other model: the WS model shows a high 
clustering but without the power-law degree distribution,
while the BA model with the scale-free nature does not possess the
high clustering. 
In this work, we propose a network model which has \textit{both}
the perfect power-law degree distribution \textit{and} the high clustering. 
Furthermore, in our model, the degree of the clustering, measured by
the clustering coefficient (see below), is shown to
be tunable and thus controllable by adjusting a parameter of the model.
  
We start from the definition of a network as a graph $\mathcal{G}=
(\mathcal{V},{\cal
E})$, where $\mathcal{V}$ is the set of vertices and $\mathcal{E}$ is
the set of edges~\cite{somegraphtheorybook}.  An edge connects pairs of 
vertices in $\mathcal{V}$ and not more than one edge may connect a specific
pair of vertices.  
To quantify the clustering, Watts and Strogatz introduced the clustering
coefficient $\gamma \equiv \langle \gamma_v \rangle$ with the
average $\langle\:\cdots\:\rangle$ for all vertices in $\mathcal{V}$.
The local clustering coefficient $\gamma_v$ for the vertex $v$ is
defined as follows: Suppose that the vertex $v$ has $k_v$ 
neighbors ($k_v$ is called the degree of the vertex $v$,
a neighbor is a vertex separated by exactly one edge).
For those $k_v$ neighbors, there can exist at most 
$\dbinom{k_v}{2} = k_v (k_v-1)/2$ edges connecting two of $k_v$ vertices.
If one defines $|\mathcal{E}(\Gamma_v)|$ as the number of
actual edges existing 
in the network connecting those neighbors, the local clustering
coefficient is written as~\cite{WS} 
\begin{equation}
\gamma_v \equiv \frac{|\mathcal{E}(\Gamma_v)|}{\dbinom{k_v}{2}}.
\end{equation}
>From the above definition, it is clear that $\gamma$ is  a measure of 
the relative number of triads (fully connected subgraphs of three vertices). 
Note also that $\gamma$ is strictly in the interval $[0,1]$ with the upper 
limit attained only for a fully connected graph. In a social acquaintance
network, for example, $\gamma=1$ if everyone in the network knows
each other. It should be noted that even though the BA model 
successfully explains the scale-free nature of many networks, it
has $\gamma \approx 0$ and thus fails to describe correctly
networks with the high clustering, such as social networks.

We below review briefly the BA model of the scale-free network and
present our model for the scale-free network with the high clustering.
The BA model~\cite{SF} is defined as follows:
\begin{itemize}
\item Initial condition: To start with, the network consists of
  $m_0$ vertices and no edges.
\item Growth: One vertex $v$ with $m$ edges is added at every time step.
  Time $t$ is identified as the number of time steps.
\item Preferential attachment (PA): Each edge of $v$ is then
  attached to an existing vertex with the probability proportional to its 
  degree, i.e.\ the probability for a vertex $w$ to be attached to $v$
  is~\footnote{In practice one may use
    $P_w = (k_w + 1) / \sum_{v\in \mathcal{V}}(k_v + 1)$, to make it possible
    for disconnected vertices to be connected. This does not change
    the resultant network geometry significantly.}
  \begin{equation}
    P_w = \frac{k_w}{\sum_{v\in \mathcal{V}}k_v} .
  \end{equation}
\end{itemize}
In the BA model, the growth step is then iterated $N=|\mathcal{V}|$ times, 
and for each growth step the PA step is iterated $m$ times for $m$ edges
of the newly added vertex $v$.  

\begin{figure}
  \resizebox*{7.5cm}{!}{\includegraphics{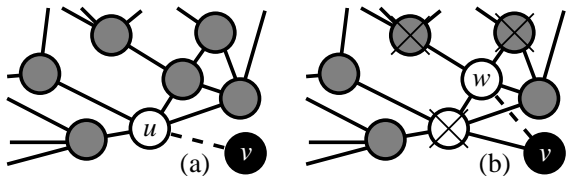}}
  \caption{\textit{Preferential attachment and triad formation.}
    In the preferential attachment step (a) the new vertex $v$ chooses
    a vertex $u$ to attach to with a probability proportional to its
    degree. In the triad formation step (b) the new vertex $v$
    chooses a vertex $w$ in the neighborhood of the one linked
    to in the previous
    preferential attachment step. 
    $\times$ symbolizes ``not-allowed to attach to''
    (either since no triad would be formed, or that
    an edge already exists).}
  \label{attachment}
\end{figure}

In order to incorporate the high clustering we modify the above
BA algorithm by adding an additional step:
\begin{itemize}
  \item Triad formation (TF):  
  If an edge between $v$ and $w$ was added in the previous PA step,
  then add one more edge from $v$ to a randomly chosen neighbor of $w$.
  If there remains no pair to connect, i.e., if all neighbors of $w$ were
  already connected to $v$, do a PA step instead.
\end{itemize}
When a vertex $v$ with $m$ edges is added to the existing network, 
we first perform one PA step, and then perform a TF step with
the probability $P_t$ or a PA step with the probability $1-P_t$.
The average number $m_t$ of the TF trials per added vertex is then 
given by $m_t = (m-1)P_t$, which we take as the control parameter
in our model  (see Fig.~\ref{attachment}). It should be noted
that our model reduces to the original BA model when $m_t = 0$.

  The standard scale free network model not only generates networks with
certain geometrical properties, it suggests a mechanism for the emergence of
power-law degree distributions in evolving networks: New actors
(vertices) in a social context prefers to attach to more connected (``well
known'') actors.  The sociological interpretation for the triad formation step
is that after being acquainted with (linked to) $w$ an actor $v$ is likely
to be acquainted to $w$'s acquaintances as well.
This mechanism of the emergence of clustering is well-known, and was discussed
under the name ``sibling bias'' already in Ref.~\cite{FRO}.
Recently, Ref.~\cite{coauthorship_networks} provided empirical evidence for
both the mechanisms of triad formation and preferential attachment used in our
construction algorithm.

\begin{figure}  
\resizebox*{8.5cm}{!}{\includegraphics{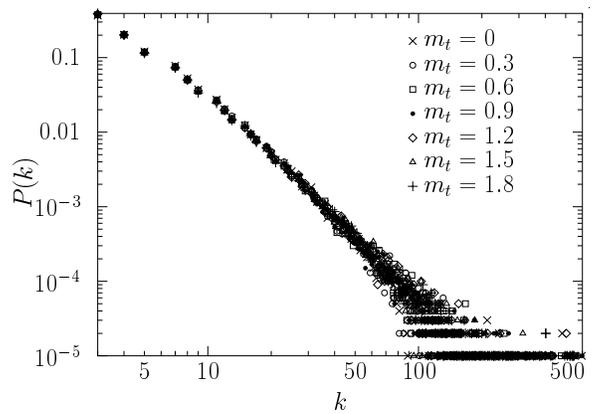}}
\caption{Degree distribution for the scale-free network model
  with tunable clustering with parameter values $m=m_0=3$, $N=10^5$
  at various values of $m_t$: At any value of $m_t$, which determines
  the average number of triad formations, $P(k)$ exhibits a power-law
  behavior like the BA model corresponding to $m_t = 0$.}
\label{c}
\end{figure}

The clustered scale-free network algorithm defined above gives the same
degree distribution as the standard scale free network, at least
if every TF step follows a PA step. 
To see this, first observe that in a PA step an
arbitrary vertex $v$ increases its degree with the rate
\begin{equation} 
\frac{\Delta k_v}{\Delta t}= A \frac{k_v}{\sum_{w\in \mathcal{V}}
  k_w}\label{pa} 
\mbox{~ ~ for a PA step},
\end{equation}
where the normalization factor $A$ for one edge is determined to be unity 
following Ref.~\cite{SF}. 
For a TF step the average increase of $k_v$ is proportional to the
probability that a vertex in the neighborhood $w$ is linked in the PA step
before, times the inverse of that vertex's degree (the probability that
$v$ is linked from $w$):
\begin{equation}
  \frac{\Delta k_v}{\Delta t}=\frac{\sum_{w\in \Gamma_v}k_w(1/k_w)}
  {\sum_{w\in \mathcal{V}} k_w}= \frac{k_v}{\sum_{w\in \mathcal{V}} k_w}
  \label{tf}\mbox{~ for a TF step},
\end{equation}
where we have used the same normalization as in Eq.~(\ref{pa}) and
$\Gamma_v$ is the neighborhood of $v$ (we use that the number of vertices in
$\Gamma_v$ is $k_v$). From Eqs.~(\ref{pa}) and (\ref{tf}) 
the total rate for one time step, composed of 
$m_t$ TF steps and $m-m_t$ PA steps, is expressed as
\begin{equation}
\frac{\Delta k_v}{\Delta t}= m_t \left( \frac{k_v}{\sum_{w\in \mathcal{V}} k_w}
\right) + (m-m_t) \left( \frac{k_v}{\sum_{w\in \mathcal{V}} k_w}\right) = 
\frac{ k_v }{ 2 t}, 
\end{equation}
which has the same form as the original BA model and thus results in 
\begin{equation}
k_v\propto t^{1/2}.
\end{equation}
Consequently, the degree of an arbitrary vertex increases as 
the square root of the time, which then yield the power-law
degree distribution:
$P(k) \sim k^{-3}$~\cite{SF}.

In the above discussion we have assumed that a TF step always
follows a PA step.
If a TF step would be proceeded by another TF step the factor
$k_w(1/k_w)$ in
Eq.~(\ref{tf}) would be replaced by
$k_w(1/(k_w-1))$ which is a small correction
when $k_w$ is large (which it is likely to be by the definition of the PA
step). And thus the resulting degree distribution would not differ much
from a power-law.  In Fig.~\ref{c}, the degree distributions $P(k)$
at various values of $m_t$ are displayed and we find that at any value
of $m_t$, the distribution is well described by the power law with
the exponent $a \approx 3$ in $P(k) \sim k^{-a}$, as is expected from
the above analytic consideration.

\begin{figure}
\resizebox*{8.5cm}{!}{\includegraphics{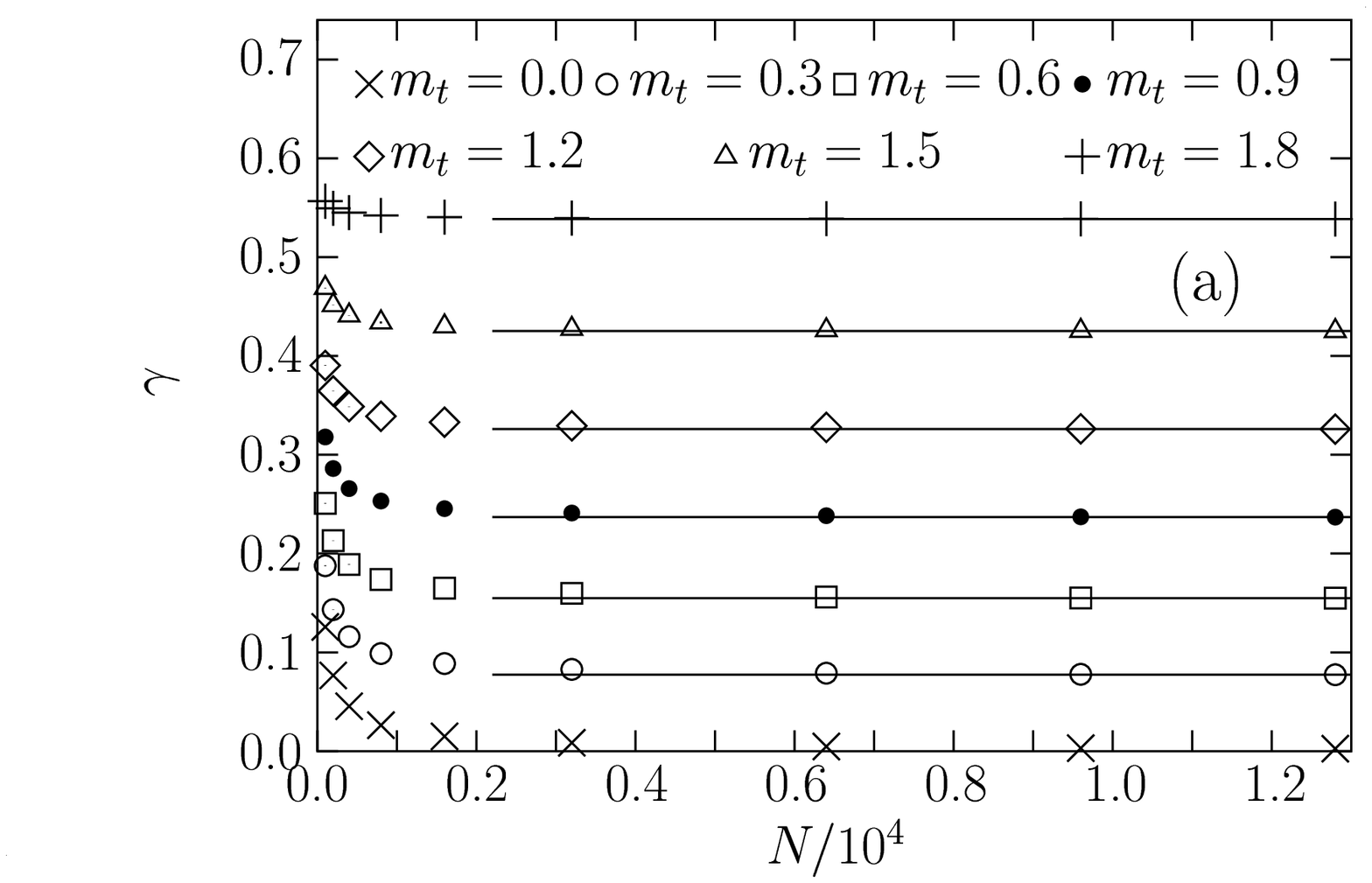}}\\
\resizebox*{8.5cm}{!}{\includegraphics{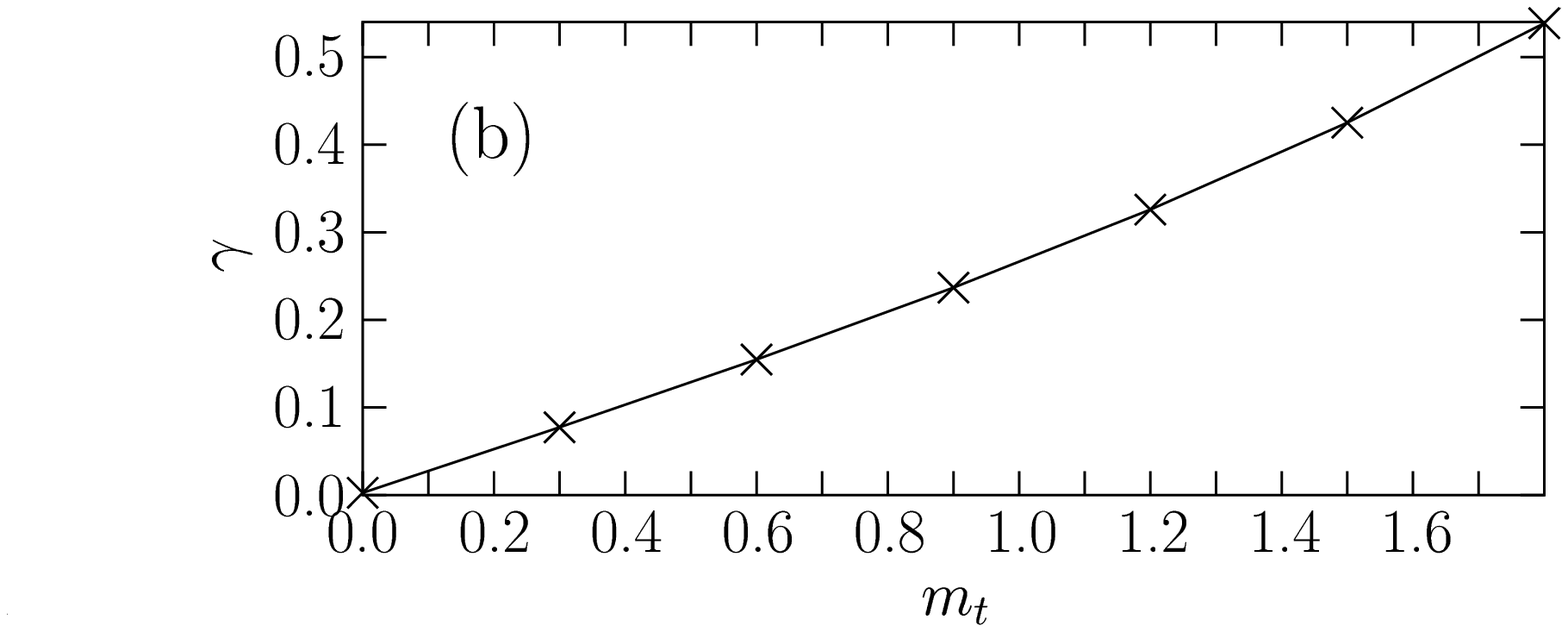}}
\caption{ (a) Clustering coefficient $\gamma$ versus the network
size $N$ at various values of the average number $m_t$ of triads
per time step. Straight lines show asymptotic values of $\gamma$ at
each $m_t$. For $m_t \neq 0$, $\gamma$ approaches a nonzero value
as $N$ is increased. (b) $\gamma (N \rightarrow \infty) $ versus $m_t$:
The clustering coefficient can be varied systematically by changing $m_t$.
}
\label{clustering}
\end{figure}

The parameter $m_t$ in our model introduces the clustering effect
into the system by allowing the formation of triads.  We only
focus on the case of $m=3$ with expectation that other values of $m$ 
should give qualitatively the same behavior.  
One expects then that for any $m_0$ a finite $m_t$ gives a 
finite clustering coefficient $\gamma$ in the thermodynamic limit
of $N\rightarrow\infty$,
whereas for $m_t=0$ (the BA scale-free network model) 
$\gamma$ goes to zero as $N$ becomes larger.  
In Fig.~\ref{clustering}(a), $\gamma$ at various values of $m_t$
is shown as a function of system size $N$. As expected, we find
that $\gamma$ approaches to a finite nonzero value as $N$
is increased at nonzero $m_t$, whereas the BA model, which corresponds
to the limiting case of $m_t = 0$ in our model, is confirmed to 
to have $\gamma = 0$.
Furthermore, we also observe
that the relation between $m_t$ and $\gamma$ is almost linear, as 
depicted in Fig.~\ref{clustering}(b). 

\begin{figure}
\vskip 0.5cm
  \resizebox*{8.5cm}{!}{\includegraphics{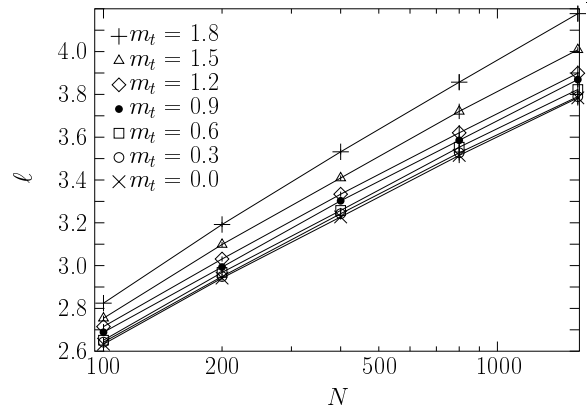}}
  \caption{The characteristic path length for the arbitrary clustered
      scale-free network model with the parameters $m=m_0=3$
      and at various values of $m_t$. Although $\ell$ becomes larger
      with $m_t$, $\ell$ is found to behave logarithmically as a
      function of $N$.}
  \label{length}
\end{figure}

>From the above observations, we conclude that our model exhibits
both the scale-free nature and the high-clustering at the same time,
while the WS model (the BA model) lacks the former (the latter) 
property. We note that in many real networks, both properties
usually coexist, and thus believe that our model is more realistic.
The triad formation step in our model, which inevitably
gives a high clustering coefficient, is expected to
make the average geodesic length smaller than the BA
network, since the edge for the triad could have been used
to connect two vertices separated by a large distance if
only the preferential attachment step was allowed.
However, the characteristic path length, defined as
the average of the geodesic length, $\ell$, is found to behave
logarithmically with the size $N$, the same behavior as
the WS model and the BA model. In Fig.~\ref{length},
we present $\ell$ versus $N$ at various values of $m_t$.
It is shown that $\ell$ becomes larger as $m_t$ is increased,
as expected. Furthermore Fig.~\ref{length} shows that the increase of
$\ell$ is logarithmic for all $m_t$.

By mimicking principles in network formation, 
a generation algorithm can construct graphs with certain topological
statistics, such as a degree distribution, clustering coefficient,
and so on. However, it should be emphasized that these kinds of algorithms
cannot claim to uniformly sample the ensemble of networks with specific
statistical properties. This drawback exists even in more general classes of
random graphs where structural biases, such as clustering, are
imposed.~\cite{markov,Str}

Recently, Klemm and Equ\'{\i}luz~\cite{klemm} have proposed a network 
model based on a finite memory of vertices, i.e., vertices become
inactive and do not get new edges after a finite number of time steps, 
and have shown 
that their growth and deactivation model 
exhibits both the high clustering and the scale-free nature.
Our model provides an alternative possibility to achieve the
same feature, the clustered scale-free nature, based on our frequent 
everyday experience on how we are acquainted by newcomers: $B$ becomes
$A$'s new friend since $B$ is introduced by one of $A$'s friends.
Even in the network of scientific citations, it is likely that
authors of paper $A$ refer paper $B$ since they have found
$B$ when they read a famous review paper $C$~\cite{citation_networks}.
This then has close resemblance to our model, the TF step accompanied
by the PA step. 
In Ref.~\cite{puniyani}, a model with both the high clustering
and the scale-free distribution has also been suggested. However,
the power-law degree distribution was assigned to the
network to start with, and the next following steps were devised
not to change the degree at each vertex. In other words,
the power-law distribution in Ref.~\cite{puniyani} was
not an emerging property in the model, which is different
from the BA model as well as our model in this work.
Very recently, we have learned about the work by Davidsen 
\textit{et al}.~\cite{davidsen},
which is based on the same observation of
triad formation as ours and has been shown to possess 
similar network properties, i.e.\ 
the high clustering, small average geodesic length, and
a scale-free distribution.
We believe, however, that our model has some advantage in describing
networks which grow in time, whereas the network model in 
Ref.~\cite{davidsen} has fixed network size.

In conclusion, we have proposed an algorithm for generation of growing 
networks
with power-law degree distribution, a logarithmic increase of the
average geodesic length, and a finite clustering. The last two properties
make the generated graphs qualify as a small-world network in the Watts
and Strogatz sense, in addition to their scale-freeness. The simple
relation between the coefficient $m_t$ and $\gamma$ further increases
the usefulness of the suggested algorithm, making it possible to tune
the clustering coefficient in a systematic way.

\begin{acknowledgments}
This work was supported in part by the Swedish Natural Research Council
through Contracts Nos.\ F 5102-659/2001 and E 5106-1643/1999.
\end{acknowledgments}

\end{document}